\documentclass[10pt]{book}

\usepackage{jfaa2e}     % jfaa 2e macros

\usepackage{latexsym}		
\usepackage{amsfonts}
\usepackage{amssymb}
\usepackage{amsthm}
\usepackage{amsmath}
\usepackage{subfigure}
\usepackage{txfonts}

\usepackage{cite}
\usepackage{hyperref}

\usepackage{ifpdf}
\ifpdf
  \usepackage[pdftex]{graphicx}
\else
  \usepackage[dvips]{graphicx}
\fi

\doi{TBA}

\newcommand{\fig}[1]{Fig.~#1}

\newcommand{\sectn}[1]{Section~#1}

\newcommand{\eg}{\mbox{\it e.g.}}
\newcommand{\ie}{\mbox{\it i.e.}}

\newcommand{\arcmin}{\ensuremath{{}^\prime}}

\newcommand{\cmb}{{CMB}}
\newcommand{\cmbtext}{{cosmic microwave background}}

\newcommand{\wmap}{{WMAP}}
\newcommand{\wmaptext}{{Wilkinson Microwave Anisotropy Probe}}

\newcommand{\isw}{{ISW}}
\newcommand{\iswtext}{{integrated Sachs-Wolfe}}

\newcommand{\lsstext}{large scale structure}

\newcommand{\nvss}{{NVSS}}
\newcommand{\nvsstext}{{NRAO VLA Sky Survey}}

\newcommand{\lcdm}{\ensuremath{\Lambda}{CDM}}
\newcommand{\lcdmtext}{\ensuremath{\Lambda} Cold Dark Matter}

\newcommand{\smhwtext}{{spherical Mexican hat wavelet}}
\newcommand{\smhw}{{SMHW}}

\newcommand{\smwtext}{{spherical Morlet wavelet}}
\newcommand{\smw}{{SMW}}

\newcommand{\snr}{{\rm SNR}}

\newcommand{\spotloc}{\mbox{\ensuremath{(l,b)=(209^\circ,-57^\circ)}}}

\newcommand{\nsigma}{\ensuremath{N_\sigma}}

\newcommand{\Den}{\ensuremath{\Omega}}
\newcommand{\Denlambda}{\ensuremath{\Den_{\Lambda}}}
\newcommand{\w}{\ensuremath{w}}

\newcommand{\eulc}{\ensuremath{\gamma}}

\newcommand{\chisqd}{\ensuremath{\chi^2}}

\begin{document}

\author{J. D. McEwen, P. Vielva, Y. Wiaux, R. B. Barreiro, \protect\break L. Cay\'on,  M. P. Hobson, A. N. Lasenby,  \protect\break E. Mart{\'\i}nez-Gonz\'alez, J. L. Sanz}

% \authorhead{J. D. McEwen}
\communicatedby{Jean-Pierre Antoine}

\chapter{Cosmological applications of a wavelet analysis on the sphere}

\footnotetext{\textit{Math Subject Classifications.}
                     --.}

\footnotetext{\textit{Keywords and Phrases.}
                     sphere, wavelets, cosmology, cosmic microwave background. }

%==============================================================================
\begin{abstract}
% We review applications of a spherical wavelet
% analysis of the \cmbtext\ (\cmb) radiation.  
The \cmbtext\ (\cmb) is a relic radiation of the Big Bang 
and as such it contains a wealth of cosmological
information.  Statistical analyses of the \cmb, in conjunction with
other cosmological observables, represent some of the most powerful
techniques available to cosmologists for placing strong constraints on
the cosmological parameters that describe the origin, content and
evolution of the Universe.  The last decade has witnessed the
introduction of wavelet analyses in cosmology and, in particular, their
application to the \cmb.  We review here spherical wavelet analyses of the \cmb\ that test the standard cosmological concordance model.  The assumption that the temperature anisotropies of the \cmb\ are a realisation of a statistically isotropic Gaussian random field on the sphere is questioned.  
Deviations from both statistical isotropy and Gaussianity are detected in the reviewed works, suggesting more exotic cosmological models may be required to explain our Universe.
We also review spherical wavelet analyses that independently provide evidence for dark energy, an exotic component of our Universe of which we know very little currently.
%
% Firstly, we review here the application of wavelets on the sphere to
% the detection of compact objects embedded in \cmb\ emissions.  This
% contamination must be removed from \cmb\ data before cosmological
% inferences may be drawn.  We then review the use of
% spherical wavelets to test the \cmb\ for deviations from Gaussianity
% and isotropy, features that may be present in non-standard
% cosmological models.  Finally, the use of spherical wavelets
% to detect the \iswtext\ effect and thus provide direct and independent
% evidence for the existence of dark energy is reviewed. 
%
The effectiveness of accounting correctly for the geometry of the sphere
in the wavelet analysis of full-sky \cmb\ data is demonstrated by the 
highly significant detections of physical processes and effects
that are made in these reviewed works.

%In this paper we review on different applications of wavelet on the
%analysis of the \emph{Cosmic Microwave Background} (CMB). This
%radiation is a relic of the past universe and it is strongly related
%with the initial matter density perturbations.  The gravitational
%interaction of those primeval matter fluctuations have generated the
%huge and complicate matter web that form the observed \emph{Large
%Scale Structure} (LSS) of the universe. The statistical analysis of
%the CMB, together with other cosmological observables, represents the
%most powerful tool for cosmologists, in order to put strong
%constraints to the cosmological parameters that defines, the origin,
%the contain and the evolution of the universe. Last decade has been
%witness of the introduction of wavelets in cosmology, and more
%specifically, in their application to the CMB. From the initial
%applications to issues like non-Gaussianity analysis and compact
%source detection, wavelets have been finding new ways to interact with
%CMB, like the cross-correlation with LSS surveys, the validation of
%the isotropy principle or the study of the topology of the universe.
%Wavelet were promising in mid 90's, anticipating interesting
%performances on incoming data sets like NASA WMAP was. They success
%has being clear, reporting some of the most interesting results on the
%WMAP data analysis on the last years. This opens a big hope for
%repeating and, perhaps, over-performing the success in the analysis of
%the incoming ESA Planck mission.
\end{abstract}
%==============================================================================

%==============================================================================
\section{Introduction}
%==============================================================================

A concordance model of modern cosmology has emerged only recently,
explaining many different observations of our Universe to very good
approximation.  
The \lcdmtext\ (\lcdm) cosmological concordance model is characterised by a universe consisting of ordinary baryonic matter, cold dark matter and dark energy (represented by the cosmological constant $\Lambda$).  Current estimates place the relative contributions of these components at 4\%, 22\% and 74\% of the energy density of the Universe respectively \cite{spergel:2006}.  Cold dark matter consists of non-relativistic, non-baryonic patricles that interact gravitationally only.  Although dark matter has yet to be observed directly, its presence has been inferred from a range of observations, such as galaxy rotation curves and the formation of the large scale structure of our Universe.  Dark energy characterises the intrinsic nature of space to expand and may represent the energy density of empty space.  Although the origin and nature of dark energy remains unclear, it may be modelled by a cosmological fluid with negative pressure, acting as a repulsive force counteracting the attractive nature of the gravitational interaction of matter.  Dark energy is required to explain the accelerating expansion of our Universe apparent from supernovae observations.  
Strong evidence supporting the \lcdm\ model is also provided by observations of the \cmbtext\ (\cmb), the relic radiation of the Big Bang.  The model not only predicts the existence of the \cmb\ but it also describes, to very high accuracy, the acoustic oscillations imprinted in the temperature fluctuations of the \cmb.

An era of precision cosmology is emerging in which many cosmological parameters of the \lcdm\ model are constrained to high precision.  Although the cosmological concordance model is well grounded both theoretically and empirically, the finer details of the model are still under much debate.  The observable predictions made by different cosmological scenarios vary only slightly, however the models themselves can have wide ranging implications for the nature of our Universe.  Due to recent high-precision data-sets and well grounded analysis techniques, precision cosmology is beginning to allow the finer cosmological details to be tested.  Independent confirmation of the general model is of particular importance also.  In this review we concentrate on analyses of the \cmb\ that apply wavelets on the sphere to test and constrain the cosmological concordance model.

\begin{figure}[t]
\centering
\includegraphics[viewport= 0 0 2048 1050,clip=,width=80mm]
    {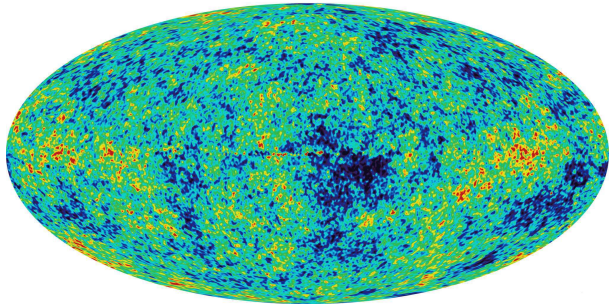}
\caption{Full-sky map of the \cmb\ anisotropies measured by \wmap\ (courtesy of the \wmap\ science team: \url{http://map.gsfc.nasa.gov/}).  The Mollweide projection is used here and subsequently to display data defined on the sphere.}
\label{fig:cmb}
\end{figure}

\cmb\ photons were emitted when the Universe was just one fifty-thousandth of its present age and since have travelled largely unhindered through our Universe, redshifting along with the expansion of the Universe.  These photons are observed today in the microwave range, exhibiting a
near-perfect black-body spectrum with a mean temperature of $2.73$K. % \cite{mather:1999}.
Observations of the \cmb\ provide a unique imprint of the early Universe.  The \cmb\ is highly isotropic, with temperature anisotropies originating from primordial perturbations at a level of $10^{-5}$ only.  However, these small temperature anisotropies contain a wealth of cosmological information.  The temperature anisotropies of the \cmb\ are thought to be a realisation of a statistically isotropic (\ie\ stationary) Gaussian random field over the sphere.  Observations of the \cmb\ are made on the celestial sphere since, 
due to the original singularity and subsequent expansion of the Universe, \cmb\ photons travel towards us from all directions.  A full-sky map of the \cmb\ temperature anisotropies measured by NASA's \wmaptext\ (\wmap) satellite \cite{hinshaw:2006} is illustrated in \fig{\ref{fig:cmb}}.  When analysing a full-sky \cmb\ map, the geometry of the sphere must be taken into account.

Wavelets are a powerful signal analysis tool owing to the simultaneous
spatial and scale localisation encoded in the analysis.  In order to
realise the potential benefits of a wavelet analysis of full-sky \cmb\
maps it is necessary to employ a wavelet analysis defined on a spherical manifold. 
Early attempts at applying wavelets to spherical \cmb\ maps were performed by \cite{barreiro:2000,cayon:2001,cayon:2003}.  In this review. however, we focus on cosmological applications that employ the wavelet transform on the sphere constructed by \cite{antoine:1999,antoine:2004}.  This methodology was derived originally entirely
from group theoretical principles.  However, in a recent work by
\cite{wiaux:2005} this formalism is reintroduced independently of the
original group theoretic formalism, in an equivalent, practical and
self-consistent approach.  
Application of this wavelet framework on the sphere for large data-sets has been prohibited previously by the computational infeasibility of any implementation. Two fast algorithms have been derived recently to perform a general directional wavelet analysis on the sphere \cite{mcewen:2006:fcswt,wiaux:2005c}. 
Without such  algorithms, cosmological analyses would not
be  computationally feasible  due to  the  large size  of current  and
forthcoming \cmb\ data-sets,  such as the 3 mega-pixel \wmap\ maps and
50 mega-pixel Planck \cite{planck:bluebook} maps respectively.
These fast algorithms and the wavelet methodology derived by \cite{wiaux:2005} are reviewed in another article of the present issue \cite{wiaux:2006:review}.

A number of analyses of the  \cmb\  have  been  performed using wavelets on the sphere to test and
constrain the cosmological concordance model that describes our Universe. The importance of accounting correctly for the geometry of  the  sphere  is  demonstrated  in these  analyses  by  the  highly statistically significant  detections of  physical  processes that  have been  made.
In this review we discuss a selection of these analyses.  In \sectn{\ref{sec:gaussianity}} we review works that test the hypothesis that the \cmb\ temperature anisotropies are indeed a realisation of a Gaussian random field on the sphere.  These works highlight slight deviations from Gaussianity and flag localised non-Gaussian features in the \cmb, thereby providing evidence for non-standard cosmological models.  In \sectn{\ref{sec:isotropy}} we review work to test the statistical isotropy (\ie\ stationarity) of the \cmb\ temperature anisotropies, another assumption of the standard cosmological concordance model.  Unexplained deviations from isotropy are detected, suggesting the need for possible alterations to the standard cosmological model.  In the final cosmological application of a wavelet analysis on the sphere that we review, we focus on verifying the existence of dark energy in the concordance model.  In \sectn{\ref{sec:isw}} we review works that use an independent method to provide evidence for dark energy and constrain its relative abundance in the Universe.  Concluding remarks are made in
\sectn{\ref{sec:conclusions}}.

%==============================================================================
\section{Gaussianity of the \cmb}
\label{sec:gaussianity}
%==============================================================================

A range of primordial processes may imprint signatures on the
temperature fluctuations of the \cmb.  The currently favoured
cosmological concordance model is based on the assumption of initial
fluctuations generated by inflation.  In the simplest inflationary
models, primordial perturbations seed Gaussian temperature
fluctuations in the \cmb\ that are statistically isotropic over the
sky.  However, this is not necessarily the case in, for example, non-standard
inflationary models or various cosmic defect scenarios.  Moreover,
non-Gaussianity may also be introduced in observational data by secondary physical effects,  residual foregrounds and instrumental systematics.
Testing whether the temperature anisotropies of the \cmb\ are a realisation of a Gaussian random field is of considerable interest.  Departures from this assumption would either provide evidence for competing scenarios of the early Universe, highlight important secondary sources of non-Gaussianity or highlight the presence of spurious signals due to foregrounds or systematic effects.

The assumption of Gaussianity has been questioned
recently with many works detecting deviations from Gaussianity in
the \wmap\ data (for a review see \cite{martinez-gonzalez:2006}).
Although the departures from Gaussianity detected in the
\wmap\ data may simply highlight unremoved foreground contamination or
other systematics in the data, which themselves are of importance for cosmological
inferences drawn from the data, if the source of these detections is
of cosmological origin then this would have important implications for
the standard cosmological model.

Ideally, one would like to localise any detected non-Gaussian
regions of the \cmb\ on the sky, in particular to determine whether they correspond
to secondary effects or systematics.  The ability to probe different
scales is also important to ensure that non-Gaussian sources present
only on certain scales are not concealed by the predominant
Gaussianity of other scales.  Wavelet techniques are thus a perfect
candidate for \cmb\ non-Gaussianity analysis.  
% 
% Planar wavelets were first used to test the Gaussianity of \cmb\
% signals on flat patches \cite{hobson:1999,pando:1998,mukherjee:2000,barreiro:2001}.  Spherical
% wavelets have since been applied to probe the full-sky \cobetext\
% (\cobe) data \cite{smoot:1992} for deviations from Gaussianity
% \cite{barreiro:2000,cayon:2001,cayon:2003},\footnote{A comparison of the performance of
% two spherical wavelets for non-Gaussianity detection has also been
% performed by \cite{martinez-gonzalez:2002}.} although no intrinsic deviations
% from Gaussianity have been reliably detected in the \cobe\ data using
% wavelets or other methods.  However, a number
% of significant detections of non-Gaussianity have been made in the
% higher resolution \wmap\ data, 
% 
Many detections of deviations from Gaussianity have been made in the \wmap\ data, some of the most significant of which have been obtained using spherical wavelets
\cite{cayon:2005,vielva:2004,mw:2004,mcewen:2005:ng,cruz:2005,cruz:2006a,cruz:2006b,mcewen:2006:ng}.
Moreover, the wavelet analysis has allowed those regions that are most
likely to induce non-Gaussianity to be localised on the sky.  In the
remainder of this section we review these analyses, focusing on the
work performed in \cite{vielva:2004,mcewen:2005:ng,mcewen:2006:ng}.  We
review the analysis procedure performed, the deviations from
Gaussianity observed and the localised non-Gaussian regions detected.

%==============================================================================
\subsection{Analysis procedure}

The wavelet transform is a linear operation.  Consequently, the wavelet
coefficients of a Gaussian map will also follow a Gaussian
distribution.  One may therefore probe a full-sky \cmb\ map for
non-Gaussianity simply by looking for deviations from Gaussianity in
the distribution of its spherical wavelet coefficients. 
Sky-cuts that would otherwise introduce non-Gaussianity may be easily accounted for by excluding those coefficients corresponding to wavelets that significantly overlap the mask.
The analysis procedure consists of first taking the spherical wavelet transform of
the \wmap\ data at a range of scales.  
Wavelet dilation scales of $14\arcmin$--$1000\arcmin$ are considered in \cite{vielva:2004}, whereas in \cite{mcewen:2005:ng,mcewen:2006:ng} the analysis is focused on the scales where non-Gaussianity was detected by \cite{vielva:2004}, \ie\ on wavelet scales ranging from $50\arcmin$--$600\arcmin$.  The effective size of a wavelet on the sky is approximately twice the wavelet dilation.  Five evenly spaced azimuthal \eulc\ orientations in the domain $[0,\pi)$ are considered for the directional wavelets applied in \cite{mcewen:2005:ng,mcewen:2006:ng}.\footnote{Identical orientations are considered at each point in the analysis performed by \cite{mcewen:2005:ng,mcewen:2006:ng}, in contrast to a continuous orientational analysis that could be performed using steerable wavelets \cite{wiaux:2006:review}.  The latter approach is the focus of current work.}
Skewness and excess kurtosis test statistics are calculated from the wavelet
coefficients in order to examine the Gaussianity of the data.  
For a Gaussian distribution skewness and excess kurtosis are zero, hence any
deviation from zero in these wavelet statistics, on any
particular scale or orientation, highlights possible deviations from
Gaussianity in the \cmb\ temperature anisotropies.  Monte Carlo simulations are performed to construct
significance measures for any candidate deviations from Gaussianity.
Gaussian simulations of \cmb\ data are constructed that model
carefully the \wmap\ observing strategy.  An identical analysis
to that performed on the \wmap\ data is applied to the simulated \cmb\ maps, in order to constrain the
significance of any detections of non-Gaussianity made from the data.
Positive detections of deviations from Gaussianity are then used to
localise regions on the sky that are the most likely sources of any
non-Gaussian signal observed.

%==============================================================================
\subsection{Results and discussion}

We review here the detections of non-Gaussianity made using the \smhwtext\ (\smhw)
and the directional \smwtext\ (\smw) (see \cite{wiaux:2006:review} for definitions and illustrations of these wavelets).
The test statistics corresponding to the most significant detections of deviations from Gaussianity are displayed in \fig{\ref{fig:stat_plot}}, with confidence intervals constructed from Monte Carlo simulations also shown.  Deviations from Gaussianity are observed well outside of the 99\% confidence region
for both wavelets.  The most significant detection made with the
\smhw\ occurs in the kurtosis of wavelet coefficients, whereas the
most significant detection made with the \smw\ occurs in the skewness.
On examining the statistical significance of detections in more
detail, the deviations from Gaussianity are made at 99.9\% and 99.3\%
using the \smhw\ and the \smw\ respectively, using a \chisqd\ test
based on the aggregate set of skewness and kurtosis test statistics to
compute these significance levels.

A wavelet analysis inherently allows the spatial localisation of interesting signal characteristics.  
The most pronounced deviations from Gaussianity in the \wmap\ data may therefore be {loca\-lised} on
the sky.  For each wavelet, the wavelet coefficients corresponding to the most {signi\-fi\-cant}
detection of non-Gaussianity are displayed in \fig{\ref{fig:coeff_wmap1}}, accompanied by corresponding thresholded coefficient maps to localise the most pronounced deviations from Gaussianity.
To investigate the impact of these localised regions on the initial
detections of non-Gaussianity, the corresponding coefficients
are removed from the calculation of skewness and kurtosis
test statistics.  
After making this correction the non-Gaussian signals detected by each wavelet (\fig{\ref{fig:stat_plot}}) are eliminated completely.
The localised regions identified do indeed 
appear to be the source of the
non-Gaussianity detected.  
In a continuation of the work of \cite{vielva:2004}, the localised
regions detected by the \smhw\ analysis are examined in more detail in
\cite{cruz:2005,cruz:2006a,cruz:2006b,cayon:2005}.  The large cold spot at
Galactic coordinates \spotloc\ is found to be the predominant source
of non-Gaussianity detected in the kurtosis of the \smhw\
coefficients.  When this cold spot is removed the kurtosis of \smhw\
coefficients is consistent with Gaussianity.  The cold spot appears to be an anomalous region on the sky not consistent with Gaussian \cmb\ temperature anisotropies, however its origin remains unknown.  Possible origins of the cold spot remain the focus of current research.

Using a spherical wavelet analysis, the hypothesis that \wmap\ observations of the \cmb\ are a realisation of a Gaussian random field on the sphere has been rejected.  The effectiveness of the wavelet analysis on the sphere is demonstrated by the highly significant detections of deviations from Gaussianity that have been made.  The source of the detected non-Gaussian signals remains unknown.  Various tests
to examine foreground contamination and instrumental systematics indicate that
these factors are not responsible for the non-Gaussianity observed.
Although it may still be the case that the deviations from Gaussianity
observed are due to these factors or other astrophysical processes, 
if they are indeed of cosmological origin this would have profound
implications for the standard cosmological model.

\newcommand{\rotangle}{-90}

\newlength{\statplotwidth}
\setlength{\statplotwidth}{60mm}

\begin{figure}
\centering
\subfigure[\smhw\ kurtosis]
  {\includegraphics[trim=0mm 0mm -10mm 0mm,clip=,angle=\rotangle,width=\statplotwidth]{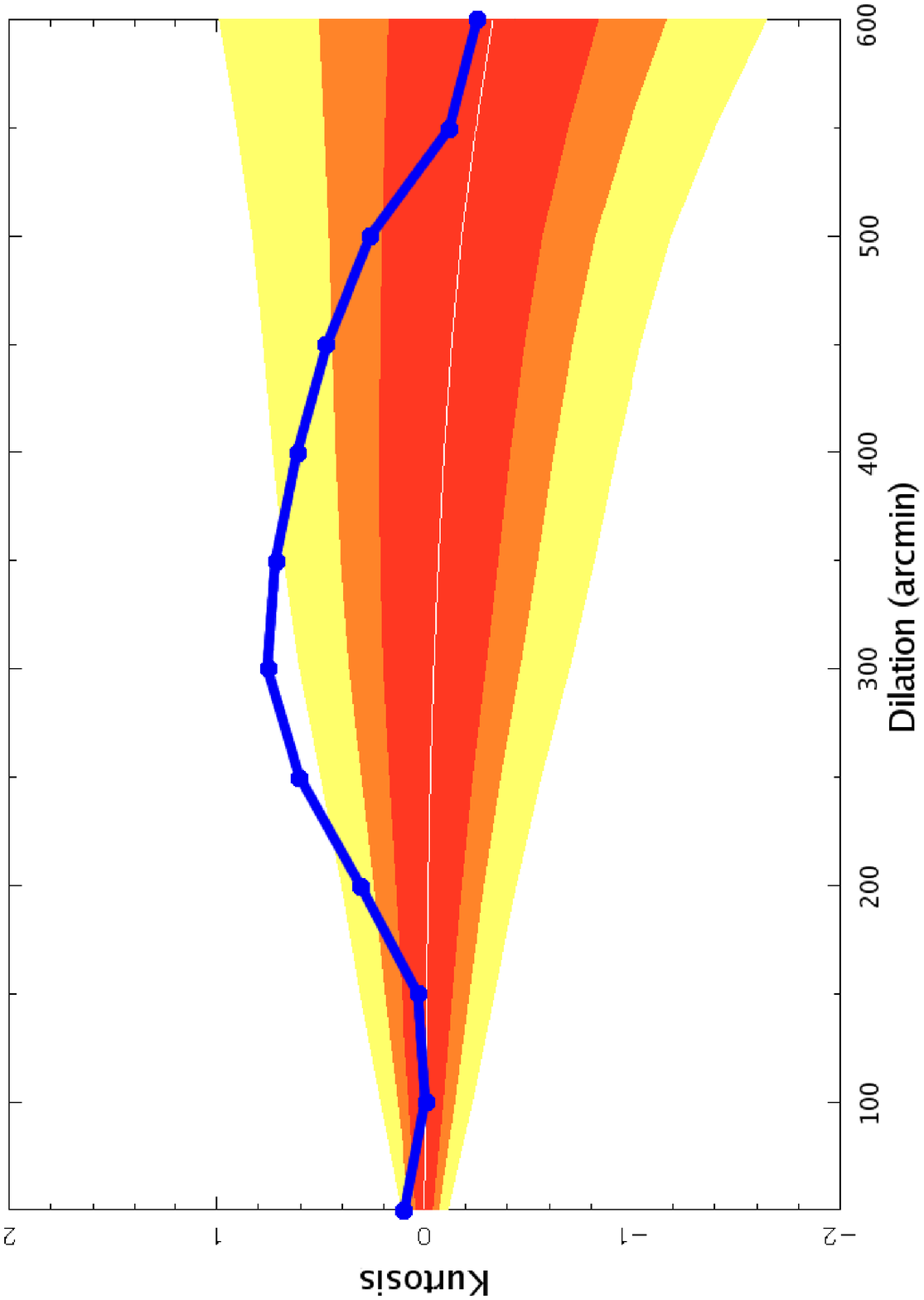}} \quad
\subfigure[\smw\ skewness (orientation \mbox{$\eulc=72^\circ$})]
  {\includegraphics[trim=0mm 0mm -10mm 0mm,clip=,angle=\rotangle,width=\statplotwidth]{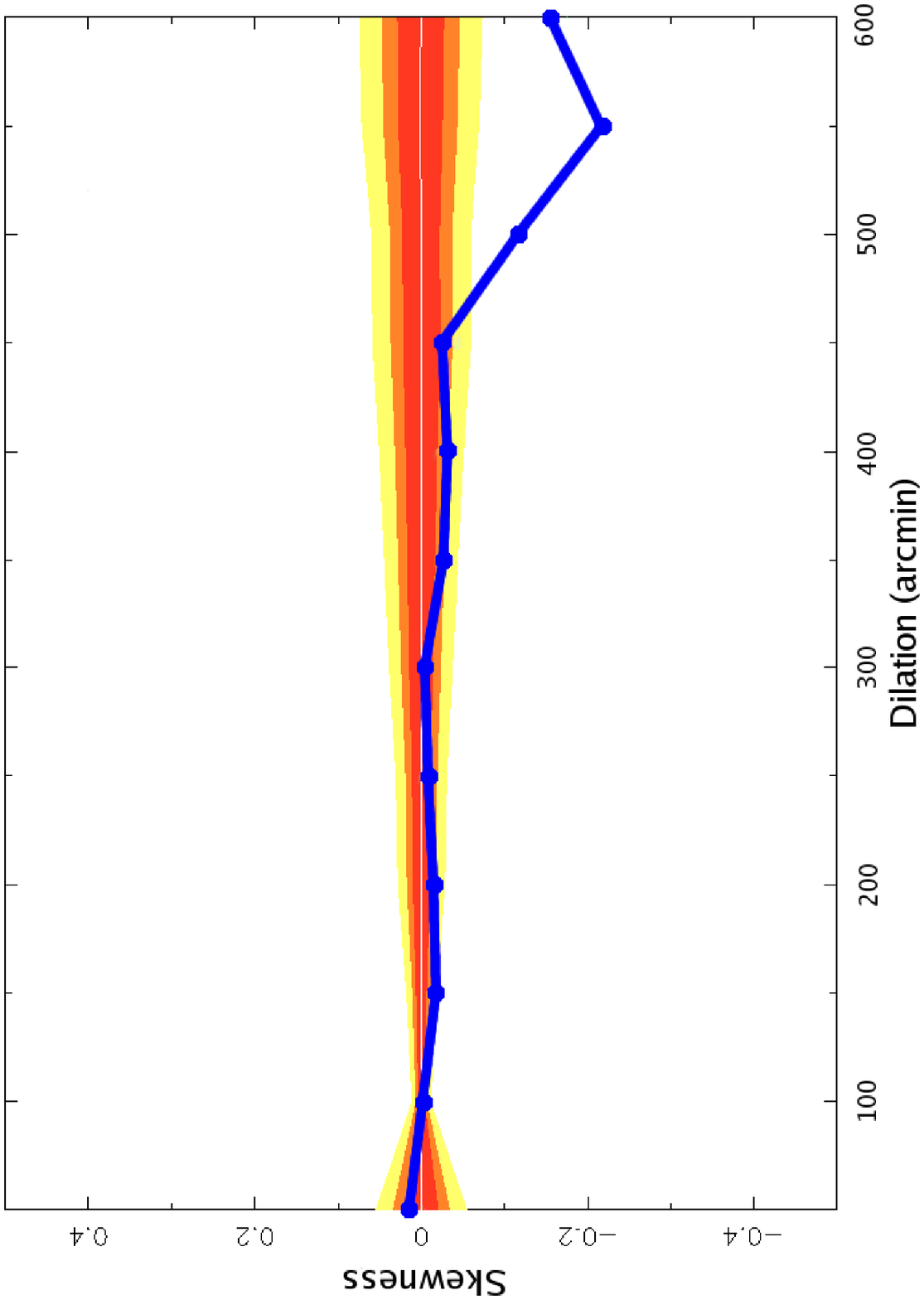}}
\caption{Spherical wavelet coefficient statistics for each wavelet.  Confidence regions
  obtained from 1000 Monte Carlo simulations are shown for 68\% (red), 95\%
  (orange) and 99\% (yellow) levels, as is the mean (solid white
  line).
  Only the orientation corresponding to the most significant deviation
  from Gaussianity is shown for the directional \smw.}
\label{fig:stat_plot}
\end{figure}

\newlength{\coeffplotwidth}
\setlength{\coeffplotwidth}{58mm}

\begin{figure}
\centering
\subfigure[\smhw\ coefficients (dilation $300\arcmin$)]
  {\includegraphics[width=\coeffplotwidth]{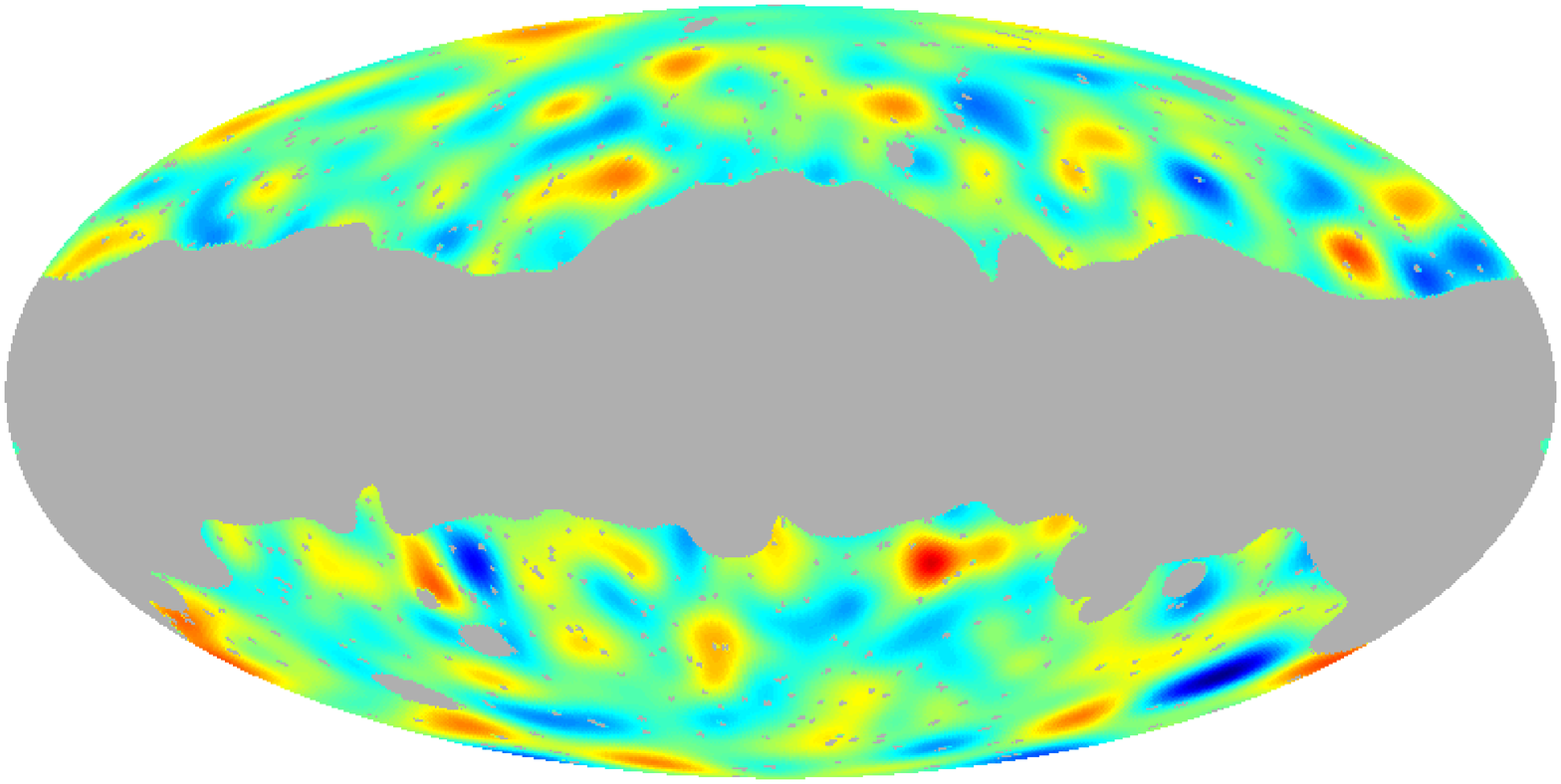} \quad
   \includegraphics[width=\coeffplotwidth]{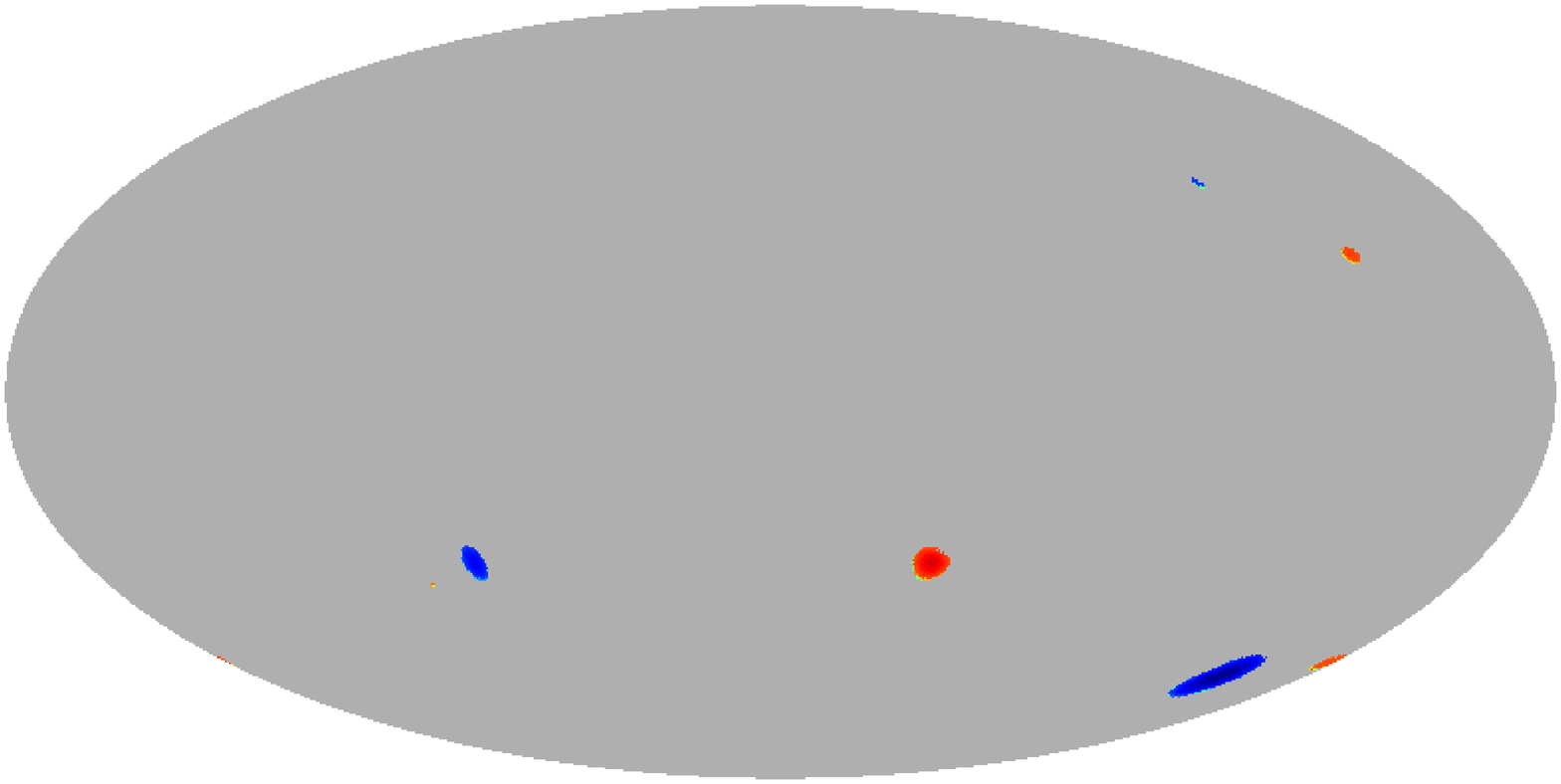} }
\subfigure[\smw\ coefficients (dilation $550\arcmin$; orientation $\eulc=72^\circ$)]
  {\includegraphics[width=\coeffplotwidth]{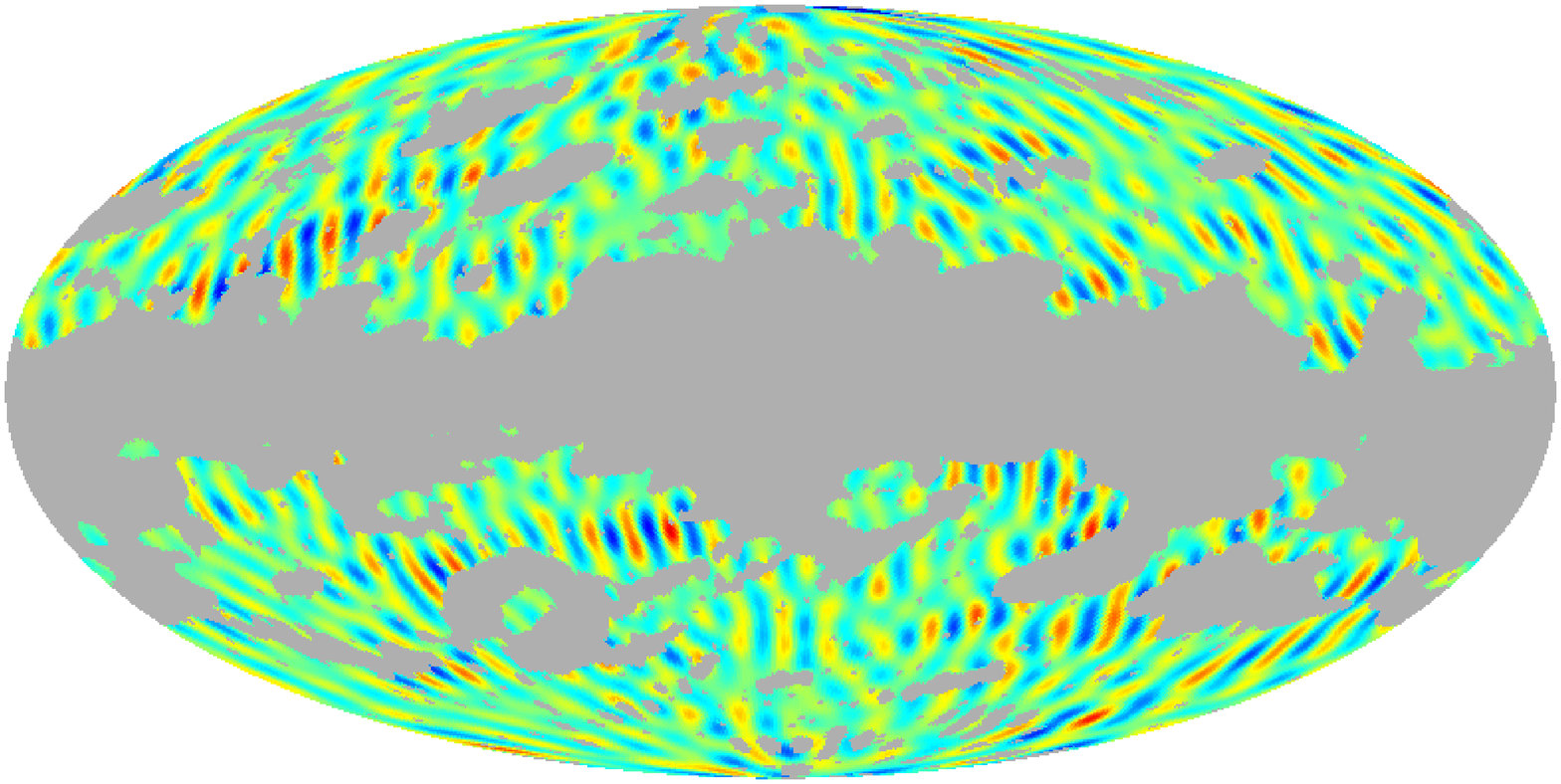} \quad
   \includegraphics[width=\coeffplotwidth]{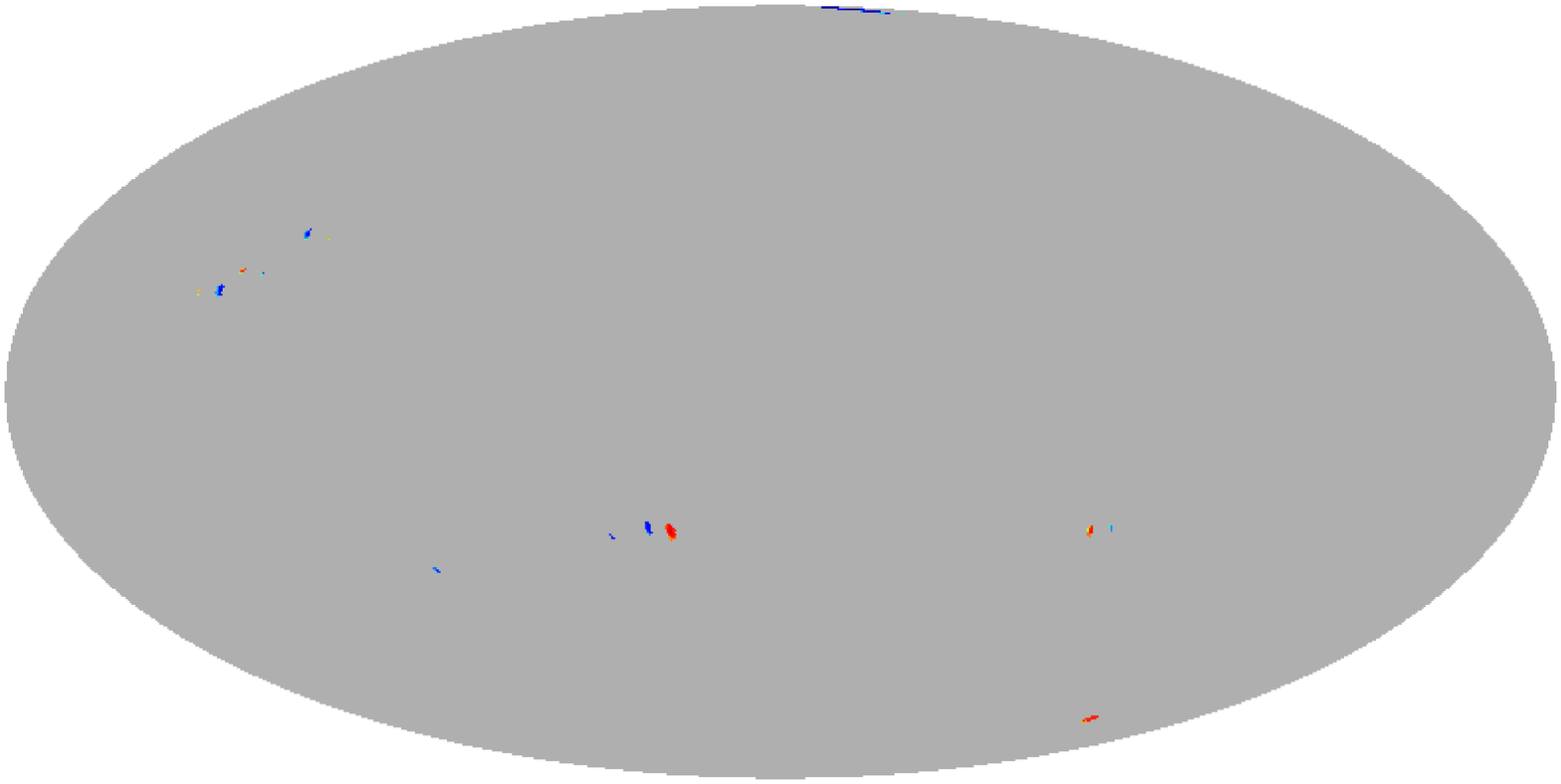} }
\caption[Spherical wavelet coefficient maps and thresholded maps]{Spherical wavelet coefficient maps (left) and thresholded maps (right).  To
  localise most likely deviations from Gaussianity on the sky, the
  coefficient maps exhibiting strong non-Gaussianity are thresholded
  so that only those coefficients above $3\sigma$ (in absolute value)
  remain.
}
\label{fig:coeff_wmap1}
\end{figure}

%==============================================================================
\section{Isotropy of the \cmb}
\label{sec:isotropy}
%==============================================================================

The statistical isotropy of the Universe is a fundamental assumption
made in cosmology, emerging from the cosmological principle.
The importance of verifying this principle cannot be overstated since
many practical cosmological calculations rely heavily upon it.  
Strong evidence in support of the cosmological principle is provided by the highly uniform temperature of the \cmb\ over the sky.  However, small deviations of the Universe from isotropy may be imprinted in the temperature anisotropies of the \cmb.  If the \cmb\ temperature anisotropies are not statistically isotropic over the sky (\ie\ statistically stationary), then this may reflect the existence of an anisotropic Universe.

Recently the isotropy of the Universe has been questioned,
with a number of works highlighting deviations from statistical isotropy in the
\wmap\ data %, while some analyses show consistency with isotropy.
(pioneered by \cite{eriksen:2005,hansen04b,land05a,copi04,deOliveira04,land05b,bielewicz05,schwarz04}). 
Possible explanations for an anisotropic Universe arise in more exotic cosmological models, such as the
so-called Bianchi models where the Universe exhibits a global shear
and rotation (see \eg\ \cite{barrow:1985,jaffe:2005,jaffe:2006b,jaffe:2006c,mcewen:2006:bianchi,bridges:2006b,cayon:2006}). 
Although some anisotropic models considered can account for some of the anomalies observed in the \wmap\ data, to date they fail to provide a consistent picture with concordance cosmology \cite{jaffe:2006b,bridges:2006b}.

There is no unique way of probing the isotropy of the Universe and a
number of groups have applied a variety of analysis techniques.  We
focus here on the alignment of local \cmb\ structures as highlighted
by the steerable wavelet analysis performed by \cite{wiaux06}. This
analysis presents to the cosmological community the first application
of steerable wavelets on the sphere (see \cite{wiaux:2005,wiaux:2006:review} for a discussion of steerable wavelets on the sphere).  We review here the
steerable wavelet analysis procedure used to probe the \wmap\ data for
deviations from statistical isotropy and discuss the deviations that are detected.

\subsection{Analysis procedure}

Under the assumption of statistical isotropy there should be no
preferred directions for the orientation of local features in the
\cmb.  A novel way of quantifying the possible alignment of \cmb\
features has been proposed by \cite{wiaux06}.  In this analysis the
number of times each direction is \emph{seen by local \cmb\ features} is
counted.  The number of times a given direction in the Universe is
seen by local features provides a unique way to determine
whether there exists any preferred direction on the sky towards which
\cmb\ structures are unexpectedly aligned.

Wavelets on the sphere are well suited for this type of
analysis, particularly steerable wavelets.  Firstly, wavelets
naturally allow one to probe different scales, hence identifying
different physical processes that may be responsible for violations of
statistical isotropy.  Secondly, steerable wavelets allow the wavelet
coefficients for any continuous orientation to be computed from a set
of basis orientations (see \cite{wiaux:2006:review}).  Using steerable
wavelets it is therefore possible to select the orientation of the most
dominant local feature.

\begin{figure}[t]
\centering
\includegraphics[width=60mm]{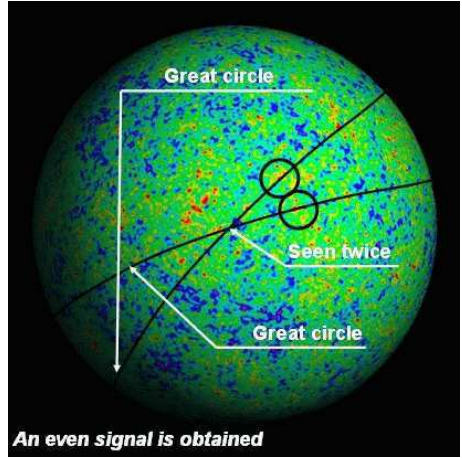}
\caption{Illustration of the steerable wavelet anisotropy analysis technique.
Pixels on the sky receive a weighted vote each time they are seen by \cmb\
features (\ie\ each time they lie on the great circle aligned with the
local feature under examination).  The votes are tallied to construct
a signal on the sphere that may be used to test the statistical isotropy of the temperature fluctuations of the \cmb.  (This figure is a modified version of that available from Max
Tegmark's web site:
\url{http://space.mit.edu/home/tegmark/index.html}.)}
\label{fig:seen}
\end{figure}

An illustration of the technique is given in \fig{\ref{fig:seen}}.
For each pixel the orientation of the local feature may be determined
using steerable wavelets as described previously.  By tracing a great
circle parallel to the local orientation, the pixels that lie on this
great circle may be selected as those regions \emph{seen} by the local
feature under examination.  Each pixel on the great circle receives a
vote proportional to the intensity the wavelet coefficient associated with the local feature.  After considering the local features at each position and
tallying the `seen' pixel votes, one recovers an even signal on the sphere that may be used to test the statistical isotropy of the \cmb.
The signal is even
(in the Cartesian coordinate system centred on the sphere) by construction 
since 
it is obtained by analysing the direction of
the local features only, without any notion about their sense (\ie\ the direction vectors are headless).
Under the
assumption of statistical isotropy, all pixels in the resultant signal
should be more-or-less `seen' to the same degree.  Simulations may be
used to quantify any departures from isotropy.

\subsection{Results and discussion}

\begin{figure}[t]
\centering
\subfigure[Anomalous probability]{\includegraphics[angle=270, width=60mm]{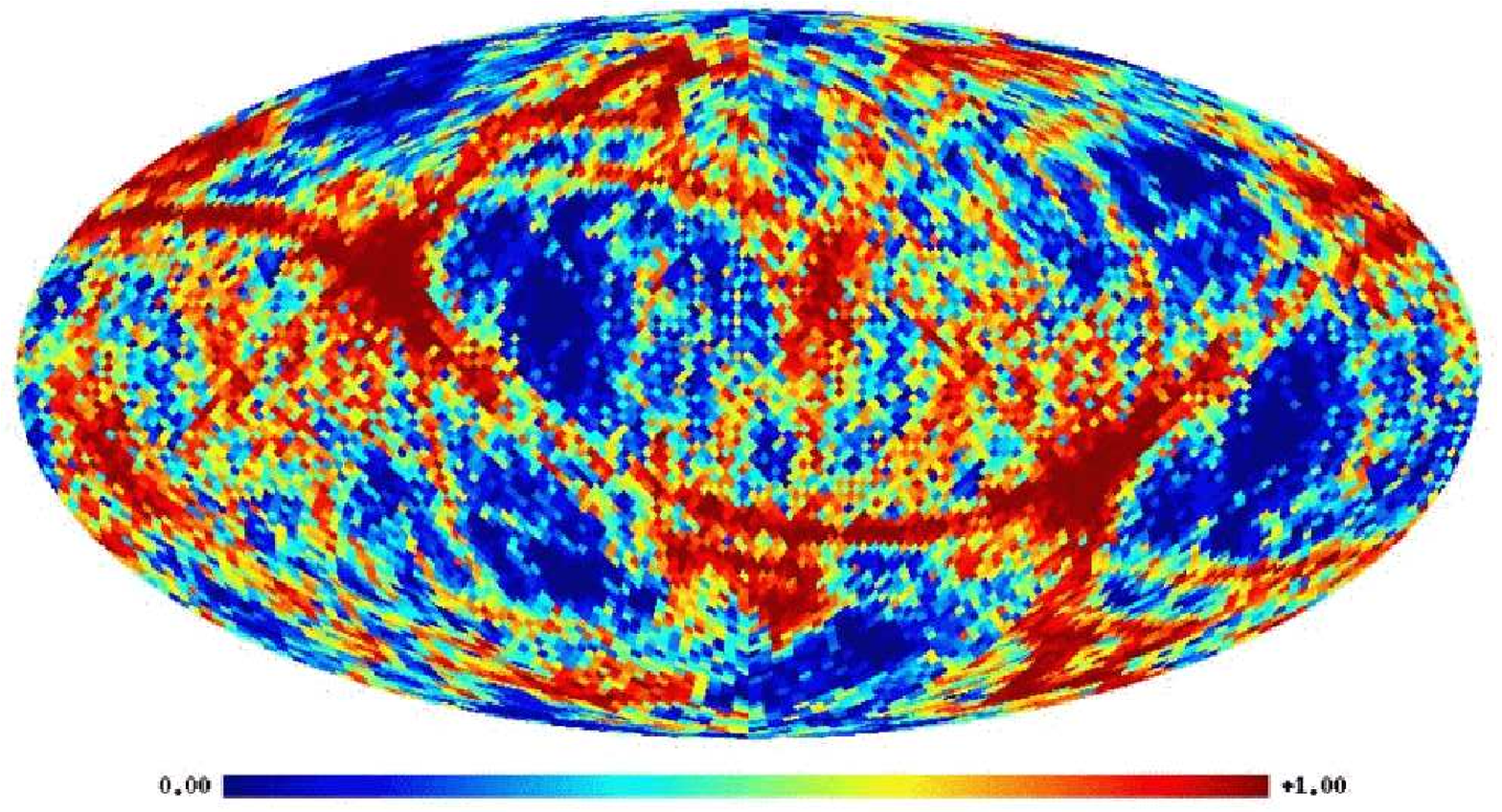}} \quad
\subfigure[Great circle defined by 20 maximally anomalous directions]{\includegraphics[angle=270,width=60mm]{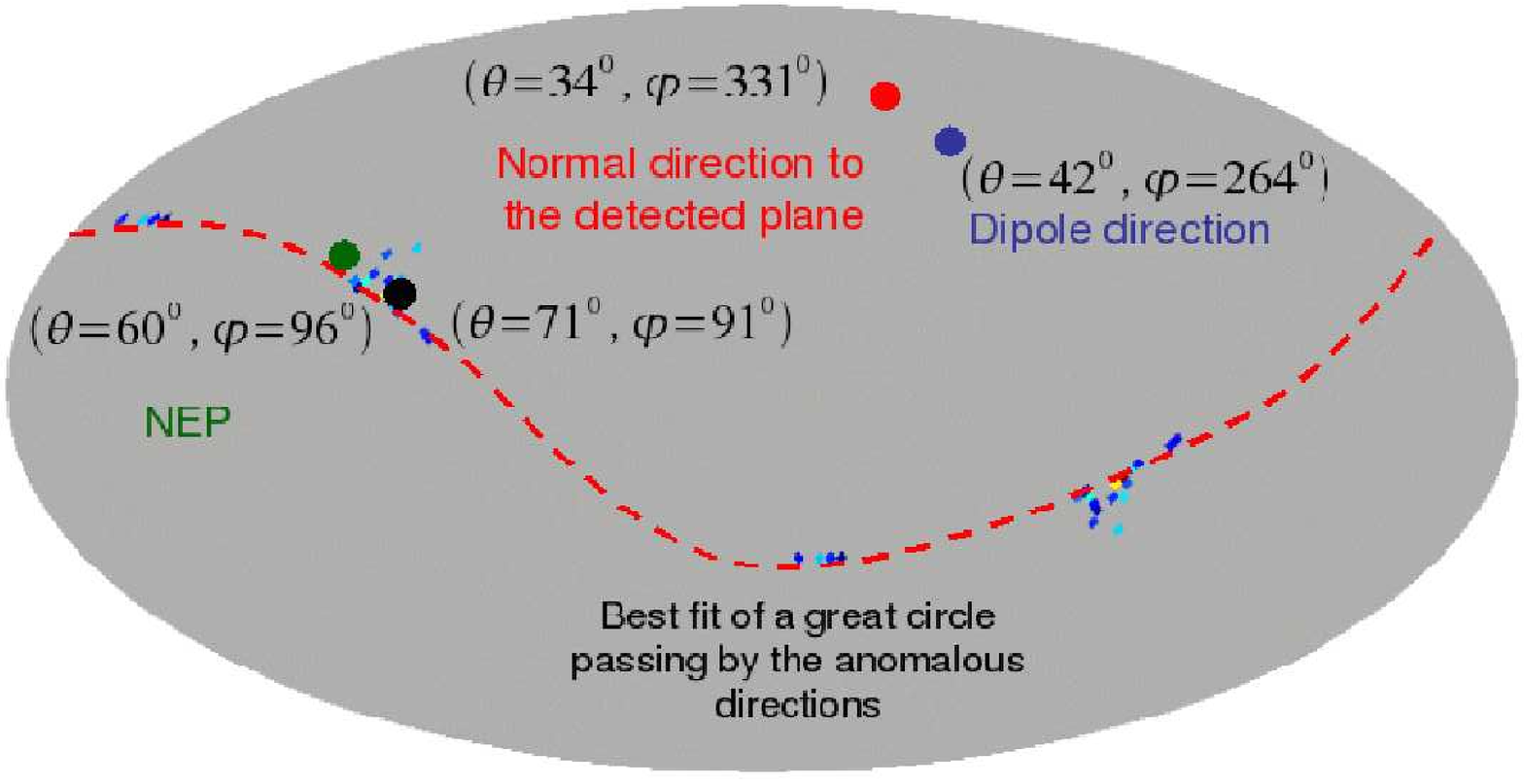}}
\caption{Anomalous directions measured in the \wmap\ data using the steerable wavelet analysis.
In panel (a) we show the probability that a given direction on
the sky is anomalous in the sense that it is seen by oriented \cmb\
features an unusually high number of times.  
In panel (b) we show the great circle defined by the 
the 20 most anomalous directions found.
% The perpendicular of this great circle lies close to the direction of the dipole and the so-called \emph{axis of evil}. The 20 maximally anomalous directions are divided in two clusters, the
% largest of which is located very close to the position of the northern
% Equatorial pole. This result synthesises the various anomalies
% reported previously by other methodologies.
}
\label{fig:anomalies}
\end{figure}

The analysis procedure described above has been applied to the \wmap\ data, indicating a highly significant violation of statistical isotropy in
the temperature fluctuations of the \cmb\ \cite{wiaux06, vielva06}.  After tallying the number of
times each pixel is seen and comparing this to simulations, it is
possible to compute the probability that a pixel receives an anomalous
number of counts, as illustrated in \fig{\ref{fig:anomalies}~(a)}.   
These results were obtained by applying
the steerable wavelet basis generated from second derivatives of a
Gaussian (see \cite{wiaux:2006:review}) at a scale corresponding to
an angular size on the sky of $8.3^\circ$.

Several great circles can be
identified from the most anomalous directions illustrated in \fig{\ref{fig:anomalies}~(a)}. In particular, one of these great circles is highlighted
by the 20 most anomalous directions (identified from those pixels for
which the probability of being anomalous is $> 99.99\%$) (see
\fig{\ref{fig:anomalies}~(b)}).
The anomalous directions are divided in two clusters, one of which
lies very close to the northern ecliptic pole.  It is interesting to note that this axis lies very close to the one corresponding to previously reported Equatorial north-south asymmetries (\eg\ \cite{eriksen:2005,hansen04b}).  In addition, the normal direction to the
great circle defined for these two clusters of anomalous directions is
close to the direction of the dipole\footnote{The direction of the
\cmb\ dipole represents the motion of the observer relative to the \cmb\
reference frame and should be removed from \cmb\ observations before
cosmological inferences are drawn from the data.} and the so-called
``axis-of-evil'' detected from low-multipole alignments
\cite{copi04, deOliveira04, schwarz04, land05b}.  The steerable
wavelet analysis has therefore synthesised the distinct anomalies
reported previously by other methodologies, suggesting a possible
interconnection.

First attempts to clarify whether this statistical anisotropy is
intrinsic to \cmb\ fluctuations or due to other sources such as
foregrounds and systematics are underway.  It has been shown by
\cite{vielva06} that there does not appear to be any frequency
dependence in the anisotropic signal detected and that the \cmb\
features themselves that are aligned towards the most anomalous
directions are homogeneously distributed on the sky.  Both of these
points seem to discard foregrounds as the source of the anisotropy.

This analysis therefore provides intriguing evidence that refutes the standard assumption that the temperature anisotropies of the \cmb\ are statistically isotropic over the sky.  Consequently, the Universe may deviate from isotropy on large scales.  These findings highlight the need to explore more exotic cosmological models that could potentially better describe our Universe.

%==============================================================================
\section{Detection of the ISW effect}
\label{sec:isw}
%==============================================================================

% The cosmological concordance model is characterised by a Universe
% containing approximately 22\% dark matter and 74\% dark energy,
% whereas the ordinary matter that we are made of constitutes 4\% of the
% Universe only \cite{spergel:2003,spergel:2006}.  Dark matter interacts
% gravitationally only and has yet to be observed directly, although
% there is other observational evidence to support its existence.  Even
% less is known about the exotic dark energy component that dominates
% our Universe.  Dark energy represents the energy density of empty
% space and is described by a cosmological fluid with negative pressure,
% acting as a repulsive force counteracting the attractive nature of the gravitational interaction of 
% matter.  

Very little is know currently about the origin and nature of dark energy, yet it dominates our Universe. 
A consistent model of dark energy in the framework of particle physics is lacking.
Indeed, attempts to predict the cosmological constant associated with
dark energy by applying quantum field theory to the vacuum energy density
arising from zero-point fluctuations predict a value that is too large
by a factor of $\sim10^{120}$.  Despite the lack of current understanding of dark
energy, the effects of a generic dark energy fluid may be modeled, describing observations of our Universe to very good approximation.  
Much of the current evidence of dark energy comes from recent
measurements of the \cmb\ \cite{hinshaw:2006} and
observations of supernovae \cite{riess:1998,perlmutter:1999}.  At this point, confirmation of the existence of
dark energy by independent physical methods is of particular interest.
We review here analyses that use the \iswtext\ (\isw) effect
\cite{sachs:1967} to independently detect and constrain dark energy.

\cmb\ photons are blue and red shifted as they fall into and out of
gravitational potential wells respectively, as they travel towards us
since emission shortly after the Big Bang.  If the gravitational potential
evolves during the photon propagation, then the blue and red shifts do
not cancel exactly and a net change in the photon energy occurs.  
The \isw\ effect is the integrated sum of these energy shifts over the photon path.
This secondary induced \cmb\ temperature anisotropy exists only in a
non-flat Universe or in the presence of dark energy
\cite{peebles:2003}. The recent \wmap\ data has imposed strong
constraints on the flatness of the Universe
\cite{spergel:2006}, hence any \isw\ signal may be
interpreted directly as a signature of dark energy.

It is difficult to isolate the contribution of the \isw\
effect from \cmb\ temperature anisotropies, hence it is not feasible to detect
the \isw\ effect solely from the \cmb.  Instead, as first proposed by
\cite{crittenden:1996}, the \isw\ effect may be detected by
cross-correlating the \cmb\ anisotropies with tracers of the local
matter distribution, such as the nearby galaxy density distribution.
A detection of large-scale positive correlation is a direct
indication of the \isw\ effect, and correspondingly, direct evidence
for dark energy.

The \isw\ effect was first detected successfully by
\cite{boughn:2004} and subsequently by many other researchers using a wide range of
different analysis techniques and data-sets.  
Furthermore, in some analyses detections of the \isw\ effect are used to constrain the dark energy parameters of cosmological models.
In these works the cross-correlation of
the \cmb\ with various tracers of the near Universe \lsstext\ is performed in
either real or harmonic space, resulting in detections of the \isw\ effect at 
the $2.5\sigma$ level approximately.  
% (\eg \cite{nolta:2004,fosalba:2003,fosalba:2004,scranton:2003,afshordietal:2004,padmanabhan:2004}).
% In all of these works the \isw\ effect is detected at the $2.5\sigma$
% level approximately.  Other works have focused on the theoretical
% detectability of the \isw\ effect for various experiments and, in some
% cases, the use of detections to constrain cosmological parameters
% \cite{afshordi:2004,hu:2004,pogosian:2004,pogosian:2005,corasaniti:2005}.
% The works discussed previously all perform the cross-correlation of
% the \cmb\ with various tracers of the near Universe \lsstext\ in
% either real or harmonic space.
% 
Recently, the cross-correlation has been evaluated in spherical
wavelet space \cite{vielva:2005,mcewen:2006:isw,pietrobon:2006}.
Since the \isw\ effect is localised to certain scales and positions on
the sky, wavelets are an ideal tool for searching for
cross-correlations due to the inherent scale and spatial localisation
encoded in the analysis.  The effectiveness of the wavelet analysis is
demonstrated by the highly significant detections of the \isw\ effect
made in wavelet space at greater than the $3\sigma$ level.  
It should be noted, however, that when information on all scales and orientations is incorporated to compute parameter constraints, the performance of the wavelet analysis is comparable to other linear techniques, as expected (assuming the fields considered are indeed Gaussian).  In the
remainder of this section we review the detections of the \isw\ effect
made by \cite{vielva:2005,mcewen:2006:isw} and the corresponding
constraints that are placed on dark energy parameters from these
detections.  
% We review the spherical wavelet covariance estimator, the
% analysis procedure and finally the results obtained and dark energy
% constraints.

%==============================================================================
\subsection{Spherical wavelet estimator}

The correlation of the wavelet coefficients may be used as an estimator
to detect any correlation between the \cmb\ and the galaxy
density distribution.  
% A positive correlation indicates a positive
% cross-correlation between the data.
%
% The wavelet coefficient correlation estimator is defined as the sum over all points on the wavelet domain sky, of the product of the wavelet coefficient maps:
% \begin{equation}
% \label{eqn:wcov}
% \wcovest^{\ndlab\tplab}(\scaleab, \eulc) = \frac{1}{N_{\eula\eulb}}
% \sum_{\eula,\eulb} \:
% \covweight_{\eula\eulb} \:
% \skywavni_\wav^\ndlab(\scaleab,\euls) \:
% \skywavni_\wav^\tplab(\scaleab,\euls)
% \spcend ,
% \end{equation}
% where $N_{\eula\eulb}$ is the number of samples in the wavelet domain sky, $\covweight_{\eula\eulb}$ is a weighting function chosen to reflect the relative size of the pixels in the wavelet domain and $\skywavni_\wav^\ndlab(\scaleab,\euls)$ and $\skywavni_\wav^\tplab(\scaleab,\euls)$ are the wavelet coefficients of the galaxy density distribution and \cmb\ respectively.  
% % 
% One may also average the correlation estimator over orientations, so that an overall correlation measure for the given scales is obtained:
% \begin{equation}
% \wcovest^{\ndlab\tplab}(\scaleab) = \frac{1}{N_{\eulc}}
% \sum_\eulc \:
% \wcovest^{\ndlab\tplab}(\scaleab, \eulc)
% \spcend ,
% \end{equation}
% where $N_{\eulc}$ is the number of samples in the wavelet domain orientational component.  This measure is still sensitive to directional structure when using a directional spherical wavelet, just as
% $\wcovest^{\ndlab\tplab}(\scaleab, \eulc)$ and
% $\wcovest^{\ndlab\tplab}(\scaleab)$
% are both sensitive to localised spatial structure in $(\eula,\eulb)$.
%
A theoretical prediction of the wavelet correlation may be specified
for a given cosmological model also.  This is derived for azimuthally
symmetric spherical wavelets by \cite{vielva:2005}.  The non-trivial
extension to directional wavelets is derived by
\cite{mcewen:2006:isw}.

% We state here the most general expression for the theoretical correlation when using directional wavelets:
% \begin{equation}
% \label{eqn:theoxcorr}
% \wcov^{\ndlab\tplab}(\scaleab, \eulc) =
% \sum_{\el=0}^\infty \:
% \clnttheo
% \sum_{\m=-\el}^\el \left| \shc{(\wav_{\scaleab})}{\el}{\m} \right|^2
% \spcend ,
% \end{equation}
% where $\shc{(\wav_{\scaleab})}{\el}{\m}$ are the spherical harmonic coefficients of the wavelet and \clnttheo\ is the theoretical cross-power spectrum of the galaxy density and \cmb\ temperature anisotropy maps.
% This result may be used to compare theoretical predictions for a given cosmological model with observations from the data, in order to place constraints of the cosmological parameters that describe the dark energy.

\begin{figure}
\centering
\includegraphics[width=60mm]{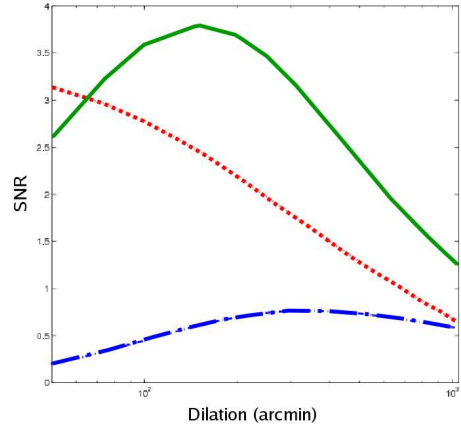}
\caption{Expected \snr\ of cross-correlation estimators for the \smhw\ estimator (solid, green), real space estimator (dotted, red) and harmonic space estimator (dot-dashed, blue).  Notice the superiority of the wavelet estimator on a large range of scales.}
\label{fig:snr}
\end{figure}

Using this result it is possible to define a theoretical expected
signal-to-noise ratio (\snr) for the wavelet correlation estimator.
Similar expected \snr s may be defined for real and harmonic space
cross-correlation estimators and are compared in \cite{vielva:2005}.
In \fig{\ref{fig:snr}} these expected \snr s are plotted for the
various techniques.  It is apparent that the wavelet
estimator is superior on a large range of scales, highlighting the
effectiveness of the spherical wavelet analysis.\footnote{A similar
comparison is performed in \cite{mcewen:2006:isw} to compare the
expected performance of various spherical wavelets.}  Of course, when
information on all scales is incorporated, the performance of the
wavelet analysis is comparable to the other linear techniques.
Nevertheless, the wavelet analysis allows one to unfold this
information and to probe only those scales with high expected \snr.

%==============================================================================
\subsection{Analysis procedure}

The analysis procedure consists of computing the wavelet correlation
estimator from the wavelet coefficients of the \wmap\ and \nvsstext\
(\nvss) data \cite{condon:1998} computed for a range of scales and
orientations.  Only those scales where the \isw\ signal is expected to
be significant are considered, that is scales
ranging over wavelet dilation $100\arcmin$--$500\arcmin$.  Anisotropic
dilations are performed in the analysis performed by \cite{mcewen:2006:isw} in order to further probe
oriented structure in the data (see \cite{mcewen:2006:fcswt} for a
description of anisotropic dilations on the sphere).  Five evenly
spaced \eulc\ orientations in the domain $[0,\pi)$ are considered.\footnote{Identical orientations are considered at each point, in contrast to the continuous orientational analysis that could be performed using steerable wavelets.  The latter approach is the focus of current work.}
Any deviation from zero in the wavelet correlation estimator for any
particular scale or orientation is an indication of a correlation
between the data and hence a possible detection of the \isw\ effect.
Monte Carlo simulations are performed to construct significance
measures on any detections made.  Gaussian simulations of \cmb\
data are constructed that model carefully the \wmap\ observing
strategy also.  An identical analysis is performed using the simulated
\cmb\ maps in place of the \wmap\ data in order to constrain the
significance of any candidate detections.  Finally, any detections of
the \isw\ effect are used to constrain dark energy parameters by
comparing observations to theoretical predictions made by various
cosmological models.

%==============================================================================
\subsection{Results and discussion}

\begin{figure}
\centering
\includegraphics[clip=,width=70mm]{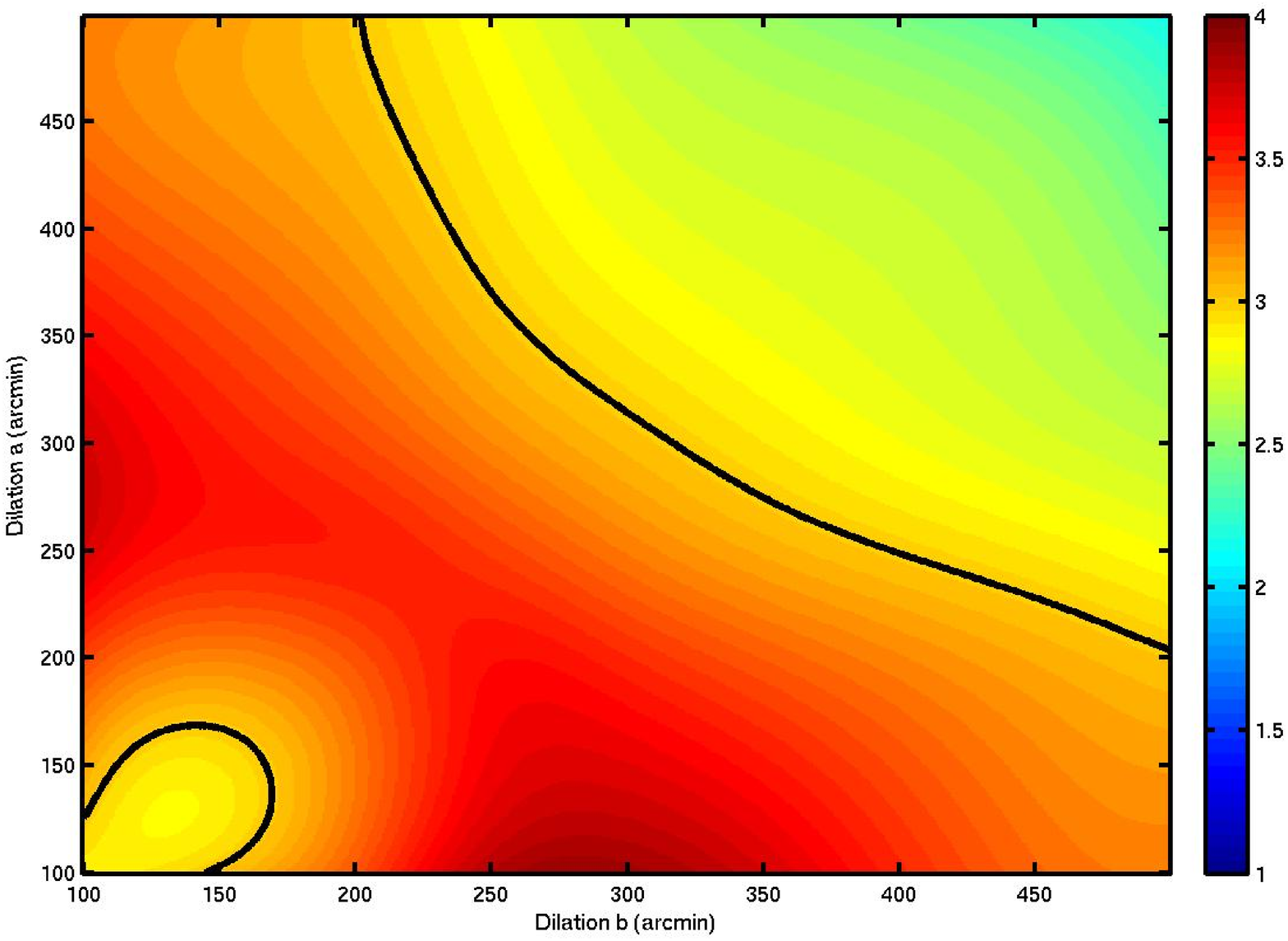}
\caption{\smhw\ correlation \nsigma\ surface.  Contours are shown for $\nsigma=3$.}
\label{fig:nsigma}
\end{figure}

We review here the results obtained using the \smhw\ only (the numerical
results obtained using other wavelets differ slightly, however the
conclusions drawn from the analysis remain the same).  A positive
wavelet correlation outside of the 99\% significance level is detected
on a number of scales and orientations using the \smhw.  On examining
the distribution of the wavelet correlation statistics from the
simulations, the correlation statistics were found to be approximately
Gaussian distributed.  This implies that the approximate significance
of any detection of a non-zero correlation can be inferred directly
from the number of standard deviations that the detections deviate by,
\ie\ from the \nsigma\ level.  In \fig{\ref{fig:nsigma}} the \nsigma\
surface is displayed in the anisotropic wavelet dilation parameter space.  The
maximum detection occurs at $\nsigma=3.9$ on wavelet scales
corresponding to approximately 10$^\circ$.

Foregrounds and systematics were analysed in detail and were
determined \emph{not} to be the source of the correlation detected.
Furthermore, the wavelet analysis inherently allows one to localise
those regions on the sky that contribute most strongly to the
cross-correlation observed.  After localising these regions they were
determined not to be the \emph{sole} source of correlation between the
\wmap\ and \nvss\ data.  This is consistent with predictions of the
\isw\ effect; namely, that one would expect to observe weak
correlations over the entire sky rather than a few strongly correlated
regions.  All tests indicate that the correlation detected in these
works is indeed due to the \isw\ effect, thus providing direct and
independent evidence for dark energy.

The positive detection of the \isw\ effect may be used to place
constraints on the properties of dark energy.  By comparing
theoretical predictions for various cosmological scenarios with
measurements made from the data it is possible to recover the
probability distribution of the data given various parameters, \ie\
the likelihood function.  The dark energy density parameter
\Denlambda\ and the equation-of-state parameter \w, which describes
the ratio of pressure to density of the dark energy fluid, are probed
in the ranges $0<\Denlambda<0.95$ and $-2<\w<0$.  The full likelihood
and marginalised distributions are illustrated in
\fig{\ref{fig:pdfall}}.  The resulting parameter estimates (see \fig{\ref{fig:pdfall}}) are consistent with
those made from numerous other analysis techniques and data-sets (\eg\
\cite{spergel:2006}).

The effectiveness of a spherical wavelet analysis to detect the \isw\
effect and thus make an independent detection of dark energy has been
demonstrated.  Although wavelets perform very well when
attempting to detect the \isw\ effect since one may probe different
scales, positions and orientations, once all information is
incorporated to compute likelihoods the performance of a wavelet
analysis is comparable to other linear techniques, as expected.
Consequently, only weak constraints may be placed on standard dark
energy parameters through the \isw\ effect, using wavelets or any
other linear technique.  Nevertheless, it is important to perform
independent tests to confirm the existence of the dark energy.
Moreover, an interesting avenue of future research is to investigate
the use of the \isw\ effect and a spherical wavelet analysis to place
tight constraints on non-standard dark energy parameters, such as the
dark energy sound speed \cite{pietrobon:2006,weller:2003}, where other
techniques have failed previously.

\newlength{\pdfwidth}
\setlength{\pdfwidth}{15mm}

\begin{figure}[t]
\centering
  \mbox{
  \subfigure[Full likelihood]{\includegraphics[height=\pdfwidth]{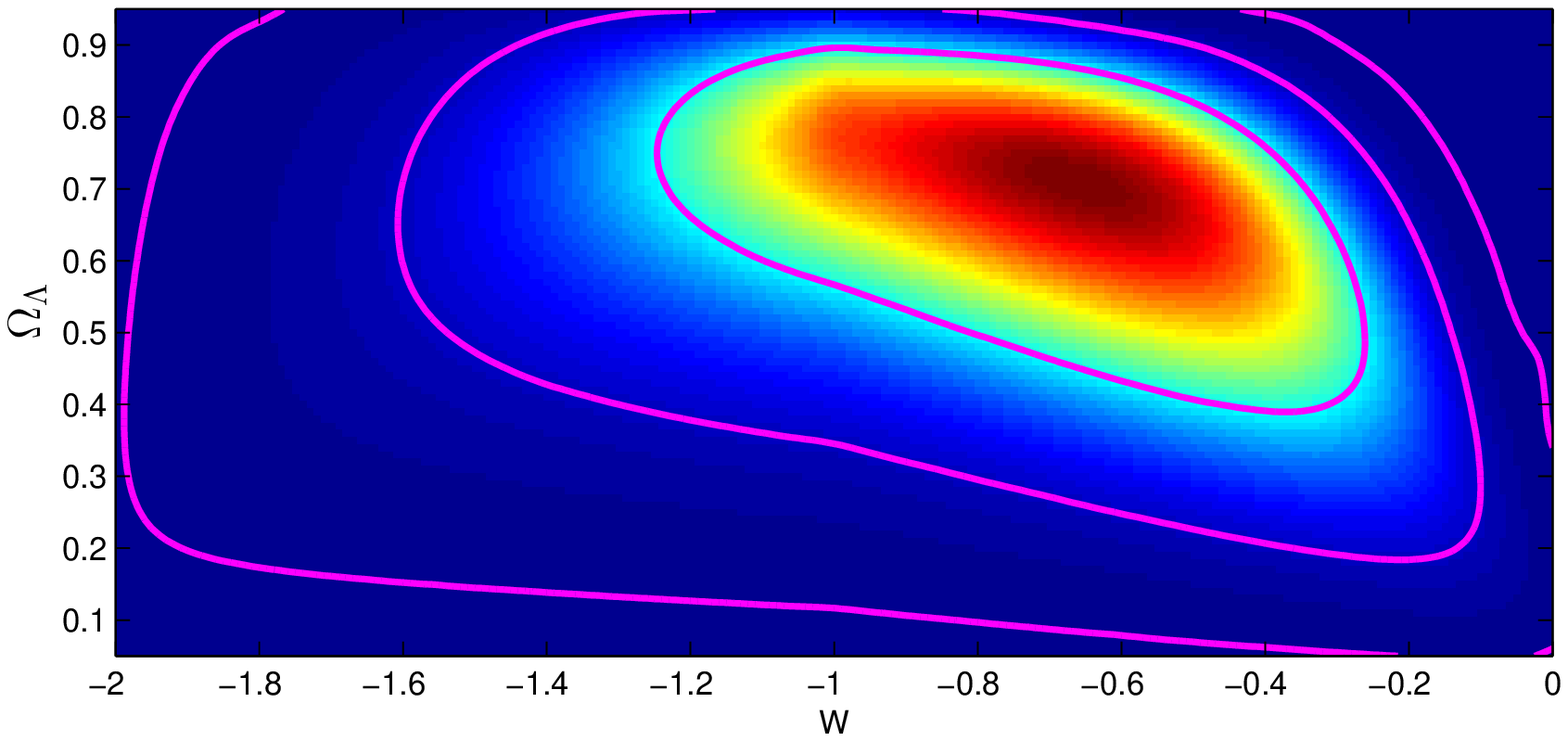}} \quad
  \subfigure[Marginalised for \Denlambda]{\includegraphics[height=\pdfwidth]{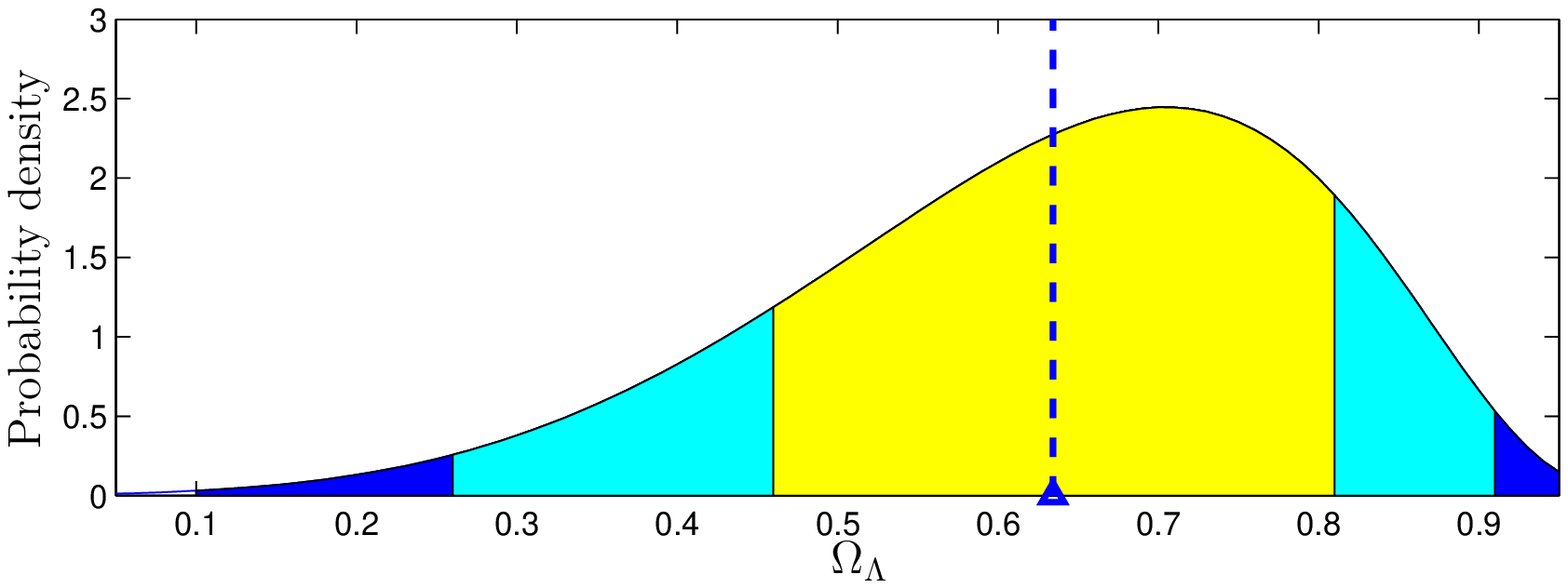}} \quad
  \subfigure[Marginalised for \w]{\includegraphics[height=\pdfwidth]{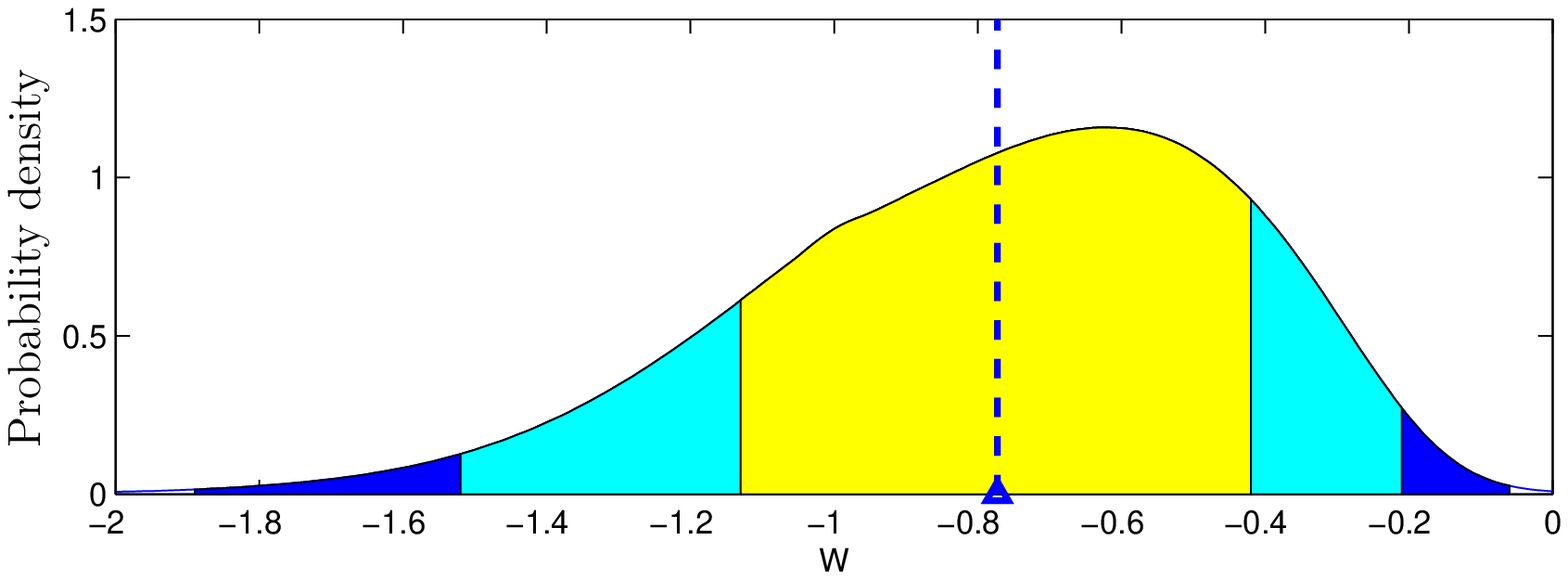}}
  }
\caption{Likelihood surfaces for parameters $(\Denlambda,\w)$.  The
full likelihood surface is shown with 68\%, 95\% and 99\% confidence
contours shown also.  Marginalised distributions for each parameter
are shown in the remaining panels, with 68\% (yellow/light-grey), 95\%
(light-blue/grey) and 99\% (dark-blue/dark-grey) confidence regions
shown also.  The parameter estimates made from the mean of the
marginalised distributions are shown by the triangle and dashed line.}
\label{fig:pdfall}
\end{figure}

%==============================================================================
\section{Concluding remarks}
\label{sec:conclusions}
%==============================================================================

In this paper we have reviewed cosmological applications of a wavelet
analysis on the sphere, concentrating on analyses of the \cmb.
The \cmb\ provides a unique imprint of the early Universe and as such contains a wealth of cosmological information; spherical wavelets may be used to extract this information.  In the analyses reviewed herein we focus on cosmological applications that adopt the wavelet transform on the sphere first proposed by \cite{antoine:1999} and revisited by \cite{antoine:2006,wiaux:2006:review}.
% , and also the transform defined by \cite{martinez-gonzalez:2002}.
We have reviewed work that has tested the assumption that the \cmb\ temperature anisotropies are a realisation of a statistically isotropic Gaussian random field on the sphere.  Highly significant deviations from both statistical isotropy and Gaussianity have been detected, suggesting 
% exotic extensions to the standard cosmological concordance model may be required.
possible extensions to the standard cosmological concordance model may be required.  No extensions can currently account for the observed anomalies, however alternative candidates should be investigated, such as non-standard inflationary scenarios, subdominant cosmic defect components or more exotic models, such as Bianchi contributions modeling a global rotation of the Universe.  Finally, we have reviewed work that independently confirms the existence of dark energy through a detection of the \isw\ effect.  
A detection of dark energy is made at a highly significant level and is used to constrain the dark energy parameters of the cosmological concordance model.
In addition to the spherical wavelet applications reviewed in detail here, wavelets on the sphere have also been used to test the global topology of the Universe \cite{rocha:2004}.  This
application presents an interesting avenue for further research.

The effectiveness of correctly accounting for the geometry of the
sphere in the wavelet analysis of full-sky \cmb\ data has been
demonstrated by the highly significant detections of physical
processes and effects that have been made in the reviewed works.
These analyses have focused on the currently available \wmap\ data, however the
forthcoming Planck data will provide full-sky \cmb\ maps of
unprecedented precision and resolution.  
% Fast wavelet analyses on the sphere will become increasingly important in this setting.
Future wavelet analyses of
Planck data may therefore outperform the already successful
analyses performed to date.
Furthermore, new and novel spherical wavelet analyses of the \cmb\ may
provide further insight in understanding the nature of our Universe.

%This paper reviews some of the cosmological applications of wavelet on
%the sphere, from the initial issues of non-Gaussianity and compact
%source detection, to most novel ones like the validation of the
%assumption of the statistical isotropy of the Universe and the
%cross-correlation of the \cmbtext (CMB) with the \lsstext (LSS).  It
%is worth to mention that wavelet have been applied to other deal with
%other cosmological analyses like the characterisation of the LSS
%(cite), denoising (cite) and the study of the topology of the Universe
%(cite).

%The application of wavelet on cosmology started in mid-90s, and, since
%then, the number of applications, papers and cosmologists interested
%on them has largely increased. For instance, around the 80\% of the
%journal papers on the issue has been published since 2000. The rest
%were so during the eight years previously to 2000.

%Validated by their success in the analysis of unprecedented CMB data
%sets like NASA WMAP, we believe that their application on future CMB
%data (like the one expected from the ESA Planck satellite) will keep
%producing outperforming results.

%==============================================================================
\section*{Acknowledgements}

JDM is supported by PPARC.
PV is supported by a I3P postdoctoral contract from the Spanish
National Research Council (CSIC).
PV, RBB, EMG and JLS are supported by the Spanish MEC project
ESP2004-07067-C03-01.
YW acknowledges support of the Swiss National Science Foundation
(SNF) under contract No. 200021-107478/1. He is also postdoctoral
researcher of the Belgian National Science Foundation (FNRS).

%==============================================================================
\section*{References}

\bibliographystyle{habbrv}
\bibliography{bibabbrv,bib,bib2}

%==============================================================================
{\footnotesize
\centerline{\rule{9pc}{.01in}}
\bigskip
\centerline{Received October 27, 2006}
\medskip
\centerline{Revision received April 20, 2007}

\medskip
\centerline{Astrophysics Group, Cavendish Laboratory, University of Cambridge,
Cambridge CB3 0HE, United Kingdom  
           }
\centerline{e-mail: mcewen@mrao.cam.ac.uk}

\pagebreak[4]

% \medskip
\centerline{Instituto de F\'isica de Cantabria (CSIC-UC), E-39005 Santander,
Spain} 
\centerline{and Astrophysics Group, Cavendish Laboratory, University of
Cambridge, CB3 0HE Cambridge, United Kingdom
           }
\centerline{e-mail: vielva@ifca.unican.es}
\medskip
\centerline{Signal Processing Institute, Ecole Polytechnique F\'ed\'erale de 
Lausanne(EPFL), CH-1015 Lausanne, Switzerland
           }
\centerline{e-mail: yves.wiaux@epfl.ch}
\medskip
\centerline{Instituto de F\'isica de Cantabria (CSIC-UC), E-39005 Santander,
Spain
           }
\centerline{e-mail: barreiro@ifca.unican.es }
\medskip
\centerline{Department of Physics, Purdue University, 525 Northwestern Avenue,
West Lafayette, IN 47907-2036, USA
           }
\centerline{e-mail: cayon@physics.purdue.edu}
\medskip
\centerline{Astrophysics Group, Cavendish Laboratory, University of Cambridge,
Cambridge CB3 0HE, United Kingdom
           }
\centerline{e-mail: mph@mrao.cam.ac.uk}
\medskip
\centerline{Astrophysics Group, Cavendish Laboratory, University of Cambridge,
Cambridge CB3 0HE, United Kingdom
           }
\centerline{e-mail: a.n.lasenby@mrao.cam.ac.uk}
\medskip
\centerline{Instituto de F\'isica de Cantabria (CSIC-UC), E-39005 Santander,
Spain
           }
\centerline{e-mail: martinez@ifca.unican.es}
\medskip
\centerline{Instituto de F\'isica de Cantabria (CSIC-UC), E-39005 Santander,
Spain
           }
\centerline{e-mail: sanz@ifca.unican.es}
\medskip

\end{document}